\documentclass[twoside,12pt]{article}
\usepackage{epsfig}

\def\Journal#1#2#3#4{{#1} {#2} (#4) #3 }

\def\NPA{{\em Nucl. Phys.} A}

\def\PRC{{\em Phys. Rev.} C}

\newcommand{\be}{\begin{equation}}
\newcommand{\ee}{\end{equation}}
\newcommand{\bea}{\begin{eqnarray}}
\newcommand{\eea}{\end{eqnarray}}

\newcommand{\mm}[1]{\mathbf{#1}}
\begin{document}

\title{ \vspace{1cm} Modifications of single-particle properties in nuclear
  matter induced by three-body forces}
\author{V.\ Som\`a$^1$ and P.\ Bo\.{z}ek$^{2,1}$\\ \\
$^1$Institute of Nuclear Physics PAN, Krak\'ow, Poland\\
$^2$Institute of Physics, Rzesz\'ow University, Rzesz\'ow, Poland}
\maketitle
\begin{abstract} 
Within the self-consistent Green's functions formalism,
we study the effects of three-body forces on the in-medium spectral function,
self-energy and effective mass of the nuclear matter constituents, analyzing the density and momentum dependence.
\end{abstract}
All microscopic calculations based on bare nucleon-nucleon (NN) potentials which
aim at a realistic description of nuclear matter should include
three-body forces (TBF). 
In Ref. \cite{sb08} within the self-consistent Green's functions formalism 
we computed macroscopic and microscopic properties of both symmetric and pure
neutron matter starting from CD-Bonn \cite{cdbonn} and Nijmegen \cite{nijm} 
potentials implemented with the Urbana TBF \cite{urb}.
Three-body forces have been included via an effective two-body interaction added to
the NN potential in the in-medium $T$-matrix, derived by closing an
outgoing nucleon line with an ingoing one in the
three-body diagram in all topologically different ways.
The calculation of the energy per particle at $T=0$ in symmetric matter
confirms the necessity of taking
three-body forces into account in order to describe correctly the saturation
behavior; the resulting neutron matter equation of state and the symmetry
energy are in agreement with the current estimations.

We briefly discuss the effects of the inclusion of TBF
on the single-particle properties.
In particular we address the momentum and density dependence of the
modifications of the spectral function $A(\mm{p},\omega)$, the effective mass
$m^{\star}(\mm{p})$, which is derived from
\begin{equation}
\frac{\partial {\omega}_p}{\partial p^2}
= \frac{1}{2 m^\star} \: ,
\hspace{1cm} \mbox{where} \hspace{1cm} 
\omega_p = \frac{p^2}{2m} + \mbox{Re} \, \Sigma (\mm{p}, \omega_p) \: ,
\end{equation}
and the self-energy at the quasiparticle pole 
$\mbox{Re} \, \Sigma(\mm{p},\omega_p)$. 

In Fig. 1 we present these modifications in the form of the ratio
between the quantities computed with and without TBF, for isospin symmetric
and pure neutron matter at zero temperature. The results are shown for the
CD-Bonn potential, and are qualitatively similar when the Nijmegen interaction
is used. 
For three different densities the ratios are displayed as a function of the 
momentum up to and above the Fermi momentum $k_F$.
Overall we observe changes of about 20-30 $\%$, the largest at higher
densities as expected from the density dependence of the three-body
potential. 

The spectral function is broadened at all densities in
symmetric matter, while the opposite effect is present in pure neutron
matter, with a narrowing of the peak.
The height of the peak can be considered in order to study
the momentum dependence of these effects.
We notice that in both cases the modification gets
stronger in the vicinity of the Fermi surface, signaling a change in the
scattering behavior, and decreases in the high momenta region where
$X^{3\mbox{\tiny{-body}}}/X^{2\mbox{\tiny{-body}}}
\approx 1$.

\begin{figure}[tb]
\label{fig:six}
\epsfysize=9.0cm
\begin{center}
\begin{minipage}[t]{17cm}
\epsfig{file=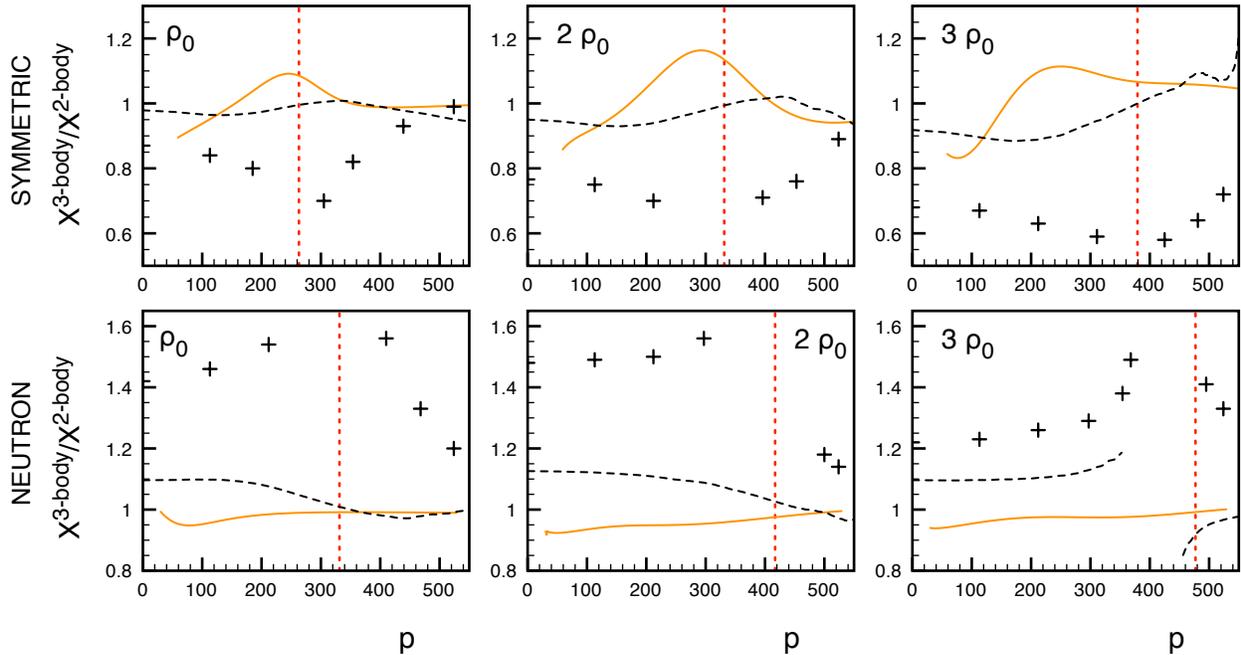,scale=0.5}
\end{minipage}
\begin{minipage}[t]{16.5 cm}
\vspace{-.8cm}
\caption{(Color online)
Ratios (with vs. without TBF) of single-particle quantities 
%$X^{3\mbox{\tiny{-body}}}/X^{2\mbox{\tiny{-body}}}$ 
as a function of the momentum.
$X$ stands for the real self-energy at the quasi-particle pole
$\mbox{Re} \, \Sigma(\mm{p},\omega_p)$ (dashed line), the effective mass
$m^{\star}(\mm{p})$ (continuous line) and the value of the peak of the
spectral function $A(\mm{p},\omega_{\mbox{\tiny{peak}}})$ (crosses). The upper
(lower) panels refer to symmetric (pure neutron) matter at densities, from
left to right, $\rho_0=0.16 \, \mbox{fm}^{-3}$,  $2\,\rho_0$ and $3\,\rho_0$.
The vertical dashed line indicates the Fermi momentum.
\vspace{-.4cm}
}
\end{minipage}
\end{center}
\end{figure}
	
The self-energy and the effective mass also show a different response depending on the isospin asymmetry. In general the real part of the self-energy is
decreased (increased) by TBF and the effective mass at the Fermi surface
is larger (smaller) when TBF are included in symmetric (neutron) matter. 
The largest modifications
appear for intermediate momenta and 
are suppressed, in both cases,  for $k>k_F$, with the ratios approaching
the value of one.  

The inclusion of three-body forces does modify the scattering behavior
of nucleons in the dense medium. 
At low and intermediate momenta all single-particle observables are affected.
Around $k_F$ the single-particle spectrum
undergoes the largest changes, then above the Fermi surface the effects
in general are suppressed. These modifications can be reflected in the
estimation of the in-medium nucleon-nucleon cross-section, relevant for
heavy-ion reactions and the physics of neutron stars.
\\
%The inclusion of TBF is necessary in order to
%obtain a realistic equation of state of the dense and hot nuclear matter.
%In this framework, the Green's functions approach allows us to address 
%the finite temperature region, currently under investigation.
\\
\textbf{Acknowledgments} Research supported in part by the Polish
Ministry of Science and Higher Education, grant No. N202 1022 33.

\end{document}